\documentclass[english]{paper}
\usepackage[T1]{fontenc}
\usepackage[latin9]{inputenc}
\usepackage[letterpaper]{geometry}
\usepackage{graphicx}

\makeatletter
\usepackage{babel}

\makeatother

\usepackage{babel}
\begin{document}

\title{{\Large The Next Frontier in UHECR Research with an Upgraded Pierre
Auger Observatory\vspace*{3mm}
 }}

\author{\emph{The Pierre Auger Collaboration}}

\maketitle

\section{Introduction}

The data taken with the Pierre Auger Observatory have led to major breakthroughs in the field of ultra-high energy cosmic rays.
Firstly, a suppression of the cosmic ray flux at energies above $5.5\times10^{19}$\,eV
has been established unambiguously \cite{spectrum}. Secondly, due to the limits from Auger
on photon and neutrino fluxes at ultra-high energy, it is now clear
that unusual ``top-down'' source processes (e.g., the
decay of super-heavy particles) cannot account for a significant part
of the observed particle flux \cite{photons,neutrinos}. Finally, there are indications of an
anisotropic distribution of the arrival directions of particles with
energies greater than $5.5\times10^{19}$\,eV \cite{agn}. These results are
typically considered as strong support of ``conventional'' source scenarios in which particle acceleration takes place at sites distributed similarly to
the matter distribution in the universe, with energy loss processes
leading to the observed flux suppression (GZK effect) and arrival
direction anisotropy. At the sources, the differential energy spectrum of injected cosmic ray particles is a power-law $E^{-\gamma}$, with
spectral index $\gamma\approx2.2$ to 2.6.

However, data on shower-development fluctuations, as well as other
composition sensitive observables that use hadronic multiparticle
production models tuned with recent LHC measurements, require consideration
of a rather different interpretation of the Auger data. One scenario
is that the observed flux suppression is indicating the upper-limit
of the capabilities of the accelerator. It may be that the uppermost end
of the cosmic ray energy spectrum is dominated by heavier nuclei
from a single source or source population, possibly within
the GZK horizon, for which the upper limit of particle acceleration
almost coincides with the energy of the GZK suppression. In contrast
to conventional expectations, models suggest that these sources would be characterized by a harder effective injection index with the
different mass components exhibiting a rigidity-dependent maximum injection
energy $E_{{\rm max}}/Z$ of a few EeV, where $Z$ is the nuclear charge.  The observed suppression of the energy spectrum mainly would stem from the source characteristics rather than being the imprint of particle propagation through the CMB. Another possibility is that there is a mixed or heavy composition
at the source and flux suppression results primarily from photodisintegration of heavy nuclei and other GZK energy losses. Whatever the acceleration
mechanism, the arrival directions of protons of sufficiently high
energy are expected to correlate with their source directions.

The data from the Pierre Auger Observatory indicate that above $10^{18.3}$ eV, UHECR primaries consist of a mix of protons together with heavier
components, with the fraction of protons becoming smaller above $10^{19}$
eV. An important caveat to this interpretation is that Auger studies
using \emph{hybrid}\footnote{Hybrid events are those with data from both the Fluorescence Detector (FD) and the Surface Detector (SD)}, as well as muon number estimates from the surface detector, show that current hadronic interaction models do not properly
describe the hadronic component of air showers. Thus, a better understanding of the particle interaction physics in UHE air showers is a crucial aspect of composition determination of UHECRs.

\section{Physics Objectives of an Upgraded Auger Observatory}

These results have motivated the Auger Collaboration to plan an upgrade
of the Auger Observatory. The primary objective of the upgrade is
\textbf{to elucidate the origin of the flux suppression}, i.e., to
differentiate a spectral cut-off due to energy loss during propagation
from a cut-off due to reaching the maximum energy achievable in the
sources. This question is both the natural evolution of the original
objective of the Pierre Auger Observatory -
to quantify the existence of a GZK-like flux suppression - and a major
step beyond it. Understanding the origin of the flux suppression will
provide fundamental constraints on the astrophysical sources. It will
resolve the question of whether some or even all UHECRs
may be of Galactic origin, and it will enable a reliable calculation
of the expected fluxes of GZK neutrinos and gamma rays, which are
sought by Auger as well as other collaborations in particle astrophysics.

The second key science objective is \textbf{to determine the mass
composition at the highest energies and to measure the flux contribution
of protons up to the highest energies.} This is an essential facet
of fully understanding the origin of the flux suppression, and critical
to understanding the nature of the sources. We aim to be sensitive
to a proton contribution as small as $\sim10$\,\%. Knowledge of
the fraction of protons is a decisive ingredient for estimating the
physics potential of existing and future cosmic ray, neutrino, and
gamma-ray detectors.

Determining the mass composition of ultra-high energy cosmic rays
is closely related to, and crucially depends upon, understanding extensive
air showers and hadronic interactions at ultra-high energy. However,
a variety of independent Auger studies using i) hybrid events, ii)
inclined showers (dominantly muonic at ground level) and iii) direct
muon number estimation from FADC traces, consistently show an excess of muons (by a factor of 1.3 to 2) compared to predictions using LHC-tuned event
generators \cite{mu-deficit}. Thus the third key science objective will be \textbf{to
understand ultra-high energy extensive air showers and hadronic multiparticle production}. With hundreds of secondary interactions being above LHC
energies in even a $10^{19}$ eV air shower, the detailed shower observations
possible with the upgraded Auger detector will enable exploration
of fundamental particle physics at energies beyond those accessible
at terrestrial accelerator facilities.
%Many types of new physics phenomena can in principle be observed or constrained, including Lorentz invariance violation, extra dimensions, micro-black hole production, and other new types of interactions above LHC energies. In this way, the upgraded Auger Observatory will potentially contribute to understanding the hierarchy problem, a major preoccupation of high energy particle physics.

\section{Technical Strategy to Achieve the Physics Objectives}

Essential to the science goals of the planned Auger upgrade is the
ability to characterize showers with much greater detail and better accuracy
than has been possible before, for as many events as possible - particularly at the highest energies. Some of the desirable features of the upgraded
detector are:
\begin{itemize}
\item the ability to measure the lateral distribution function (LDF) of
the muon and electromagnetic (EM) components separately, with better muon
identification and extending closer to the core without saturation of the SD detector signals, and

\item accurate measurement not only of the depth of shower maximum for the
EM component (as at present for hybrid events) but also the depth
at which muon production reaches its maximum, using the surface detector
alone.
\end{itemize}
There are multiple approaches to achieving the desired detector performance,
as discussed in this section. Common to all of them is the surface
detector electronics (SDE) upgrade.

The SDE upgrade is the first step in separating the muon and EM components
of the showers on an event by event basis \cite{mu-counting}. It
includes faster timing of surface detector signals,
improving significantly the ability to distinguish close-in-time
muon pulses across the entire array. The updated electronics will
also facilitate the addition of dedicated muon detectors over all
or part of the surface detector array to further improve the muon/EM
separation and reduce model-dependent systematic errors. This gives
rise to a ``boot-strap'' approach, where the model-independent,
direct muon determination abilities of the upgraded detector are used
to validate and refine the more indirect detection methods and analyses.

Distinct strategic options appear in how to use muon identification
and/or separation of the muonic and EM components of the signal, to
test and improve the hadronic physics modeling and to assign a composition-probability
to each event. The most direct method is to simply separate the muonic
and EM signals in each tank, and use timing and geometry to infer
the location of peak muon production point in the development of the air shower \cite{mpd}. A second approach is to fit the total signal in every tank, including some elements
of the FADC timing, to a superposition of templates of EM and muonic
components. This gives a more accurate energy and angular reconstruction
than the traditional method, and at the same time gives $X_{max}$\ with
remarkable accuracy
%\footnote{Depending on which technique is being used, what is reconstructed is $X_{max}$\ or $X_{max}^{\mu}$. Both observables are composition indicators; in the following we do not make the distinction between them, for simplicity.}
of about $30 \, {\rm g\, cm}^{-2}$ \cite{universality}.

The additional muon identification technologies under study and prototyping
include (1) segmentation of the interiors of surface detectors, to
separate penetrating muons from the lower-energy electromagnetic component,
and (2) placement of external particle detectors (such as RPCs or
scintillators) with the existing Auger detectors.

It is planned to operate the upgraded Pierre Auger Observatory from
2015 to 2023, which would approximately double the data set which
will have been collected prior to implementation of the upgrade.

Parallel to upgrades aiming at higher sensitivity to the muon signal, the collaboration works also on improved detection capabilities for the electromagnetic component. This is done through two initiatives. Firstly, while not a detector upgrade, it is planned to extend fluorescence light measurements into twilight by running with reduced gain, thereby increasing the duty cycle for the highest energy showers by up to a factor of two. Secondly, a rigorous R\&D project is ongoing to investigate the feasibility of radio detection as a precision tool to study the electromagnetic component of air showers.

\section{Impact of Upgrade}

$X_{\rm max}$ is an important composition indicator and thus its accurate measurement is a powerful tool to enhance anisotropy searches. The $X_{{\rm max}}$ resolution currently achieved by Auger in hybrid events is $\approx20{\rm \, g\, cm^{-2}}$. At present, the data suggest a gradual transition from light to heavier composition as the energy increases from $10^{18}$ to $10^{19.5}$ eV, with a relatively narrow range of masses present at any given energy. Therefore, good resolution on $X_{{\rm max}}$ is important for composition studies.

Only a fraction of events are seen with the fluorescence detector
which observes $X_{{\rm max}}$ directly, so a key focus of the Auger
upgrade design is to obtain good resolution on $X_{\rm max}$ and other
composition-dependent observables with the SD alone. This will give an order of magnitude increase in the statistics for events with well-determined $X_{\rm max}$, and allow composition measurements to be extended to trans-GZK energies.  Furthermore, having a well-measured $X_{\rm max}$ means that the composition of an event can be probabilistically assigned, so when this is possible with the SD it will mean that a large fraction of events can be ``composition-tagged''. This
will enable composition fractions
to be constrained up to trans-GZK energies and enhance source detection
by reducing the ``background noise'' from events which are likely
to be higher-Z nuclei and more strongly deflected, allowing them to be identified and de-weighted in anisotropy analyses.

Enhanced direct muon detection and upgraded SD electronics will allow individual showers to be studied in greater detail, providing individual longitudinal profiles of muonic shower development, as
well as separate lateral profiles of the EM and muonic components
of each shower. The
detailed spatial information on muons also will strongly constrain the
hadronic interaction modeling and enable the properties of the primaries
to be characterized with much greater confidence and precision.

Auger's excellent energy resolution augmented with
composition-estimation capabilities will allow better separation
between events from a single source and ``background'' events, and thus
enhance the potential for localizing the source via ``backtracking'' through
the Galactic magnetic field.

\section{Summary}

The Auger Collaboration plans to upgrade the Auger Observatory, with
the primary science goals:
\begin{itemize}
\item elucidate the origin of the flux suppression at the highest energies;
\item measure the composition of UHECRs up to highest energies, with sufficient
resolution to detect a 10\% proton component;
\item provide composition-tagging to facilitate anisotropy studies;
\item study hadronic interactions at center-of-mass energies an order-of-magnitude
greater than at the LHC.
\end{itemize}

Studies of the data from the Pierre Auger Observatory have significantly advanced our knowledge of cosmic rays and particle physics at extreme energies.  The proposed upgrades build on the existing strengths of the Observatory.  Including composition-sensitive measurements using the surface detector alone will provide an order-of-magnitude increase in the aperture for composition-tagging.  More detailed muon measurements will shed additional light on the interactions that propel the shower development.

By taking advantage of knowledge gained in the operations
to date, the proposed upgrade will improve greatly the capabilities of an already proven technique.  These capabilities, most especially the composition sensitivity, fit well into the larger picture of the next generation of ground- and space-based cosmic ray detectors. Auger will complement the new techniques and enhance the overall scientific reach of entire enterprise.

\end{document}